\documentclass[11pt]{article}
\usepackage{graphicx,amsfonts}
\usepackage[left=2cm,right=2cm,bottom=2cm,includeheadfoot,a4paper]{geometry}
\usepackage[onehalfspacing]{setspace}

\usepackage[comma,sort&compress]{natbib}
\bibpunct{[}{]}{,}{n}{,}{,}
\newcommand{\av}{{\bf a}}
\newcommand{\kv}{{\bf k}}
\newcommand{\KV}{{\bf K}}
\newcommand{\micron}{\mu{\rm m}}

\begin{document}

\title{Nonlinear Photonic Quasicrystals for Novel Optical Devices}

\author{Alon Bahabad$^\dag$, Ron Lifshitz$^\ddag$, Noa Voloch$^\dag$,
  and Ady Arie$^{\dag}$\\ 
  $^\dag$ School of Electrical Engineering, Fleischman Faculty of
  Engineering,\\ Tel Aviv University, Tel Aviv 69978, Israel\\
  $^\ddag$ Raymond and Beverly Sackler School of Physics \& Astronomy,\\
  Tel Aviv University, Tel Aviv 69978, Israel}

\date{January 7, 2008}

\maketitle

\begin{abstract}
  Two well-known methods for the design of quasicrystal models are
  used to create novel nonlinear optical devices. These devices are
  useful for efficient three-wave mixing of several different
  processes, and therefore offer greater flexibility with respect to
  the more common periodic nonlinear photonic crystals. We demonstrate
  applications for polarization switching as well as multi-wavelength
  and multi-directional frequency doubling.  The generalized dual grid
  method is proven to be efficient for designing photonic
  quasicrystals for one-dimensional collinear devices as well as
  elaborate two-dimensional multi-directional devices.  The cut and
  project method is physically realized by sending finite-width
  optical beams at an irrational angle through a periodic
  two-dimensional nonlinear photonic crystal. This enables two
  simultaneous collinear optical processes that can be varied by
  changing the angle of the beams.
\end{abstract}

\section{Introduction}

Twenty five years have passed since the 1982 discovery of
quasicrystals~\cite{Shechtman}, and we have yet to find a satisfying
application that takes advantage of their unique combination of
physical properties~\cite{Dubois}.  Nevertheless, interesting
applications are starting to emerge that take advantage of
quasiperiodic long-range order in metamaterials, or artificially
constructed quasicrystals\footnote{We refer the reader to
  Ref.~\cite{quasidef} for a precise definition of the term
  `quasicrystal'.}.  Most applications are based on linear photonic
crystals, where quasiperiodic modulations of the index of refraction
of a material are used in order to engineer its optical response. In
particular, the fact that there are no restrictions on the order of
the rotational symmetry of a quasicrystal is used to obtain
nearly-isotropic photonic band gaps~\cite{jin99,zoorob}.  Here we
focus on metamaterials in the {\it nonlinear\/} optical domain, where
recent technological progress has enabled to modulate the second-order
nonlinear susceptibility with micron-scale resolution in various
materials, such as ferroelectrics (our focus here), semiconductors, and
polymers. In these nonlinear photonic crystals the modulation can be
achieved by planar techniques, thereby offering either one or two
dimensions for modulation.  Moreover, there are no photonic bandgaps
in these metamaterials, because the first-order susceptibility, and
hence the refractive index, remain constant. The advantage of using
quasicrystals in this case is not in their arbitrarily-high symmetry,
but rather in the fact that there is no restriction on the
combinations of wave vectors that may appear in their reciprocal
lattices (provided that the symmetry of the quasicrystal is not of
particular importance~\cite{RLlattices,RLsymmetry}).

The novel optical devices described below are based on materials that
facilitate the nonlinear interaction between light waves in the form
of three-wave mixing. These are processes in which two incoming waves
of frequencies $\omega_1$ and $\omega_2$ interact through the
quadratic dielectric tensor $\chi^{(2)}$ of the material to produce a
third wave of frequency $\omega_3=\omega_1\pm\omega_2$; or the
opposite processes in which a single wave spontaneously breaks up into
two.  Three-wave mixing is severely constrained in dispersive
materials, where $\omega(\kv)$ is not a linear function, because the
interacting photons must also conserve their total momentum. Even the
slightest wave-vector mismatch $\Delta\kv = \kv_1 \pm \kv_2 - \kv_3$
appears as an oscillating phase that averages out the outgoing wave,
giving rise to the so-called ``phase-matching problem.'' We have
recently explained how one could fully solve the most general
phase-matching problem using well-known ideas from the theory of
quasicrystals~\cite{Lifshitz_PRL2005}. The solution is based on the
idea that in crystals\footnote{We refer the reader to
  Ref.~\cite{RLcrystal} for a detailed discussion on ``What is a
  crystal?''.}, whether periodic or not, continuous translation
symmetry is broken. As a consequence, momentum conservation is
replaced by the less-restrictive conservation law of crystal momentum
whereby momentum need only be conserved to within a wave vector from
the reciprocal lattice of the crystal. The fabrication of an efficient
frequency-conversion device is therefore a matter of {\it
  reciprocal-lattice engineering}---designing an artificial crystal, from
the quadratic dielectric field of the material $\chi^{(2)}({\bf r})$,
whose reciprocal lattice contains any desired set of mismatch wave
vectors $\Delta\kv^{(j)}$, $j=1\ldots N$, required for phase matching
any arbitrary combination of $N$ three-wave mixing processes. In fact,
the field amplitude of the output beam, in each frequency-conversion
process, is linearly proportional to the amplitudes of each of the
input beams, as well as the Fourier coefficient of the relevant
mismatch wave vector~\cite{Fejer_QPM1992}.

The idea of using a one-dimensional periodic modulation of the
relevant component of the quadratic dielectric tensor, for the purpose
of phase matching a single three-wave process, was suggested already
in the early 1960's~\cite{Armstrong1962,% 
  Freund_NonlinearDiffraction1968,Bloembergen1970}, and is termed
``quasi-phase matching''. Since then this approach has been
generalized using more elaborate one-dimensional~\cite{Fejer_QPM1992,%
  Ming_THG1997,Keren_PRL2001} and
two-dimensional~\cite{Berger_NPC1998,Broderick_Hexagonal2000,%
  Bratfalean_NPQC2005,Ma_OcatgonalNPQC_APL2005} designs, but only as
{\it ad hoc\/} solutions for multiple processes.  We argue that
engineering the reciprocal lattice, of a nonlinear photonic
quasicrystal, to contain any desired set of mismatch vectors---a task
that 25 years of research in quasicrystals have taught us how to
solve---provides the most general solution for the long-standing
problem of multiple phase-matching. Here we describe a number of novel
optical devices that have actually been fabricated using these ideas,
and tested experimentally.  In Sec.~\ref{sec:dgm} we describe devices
that have been designed using the dual-grid method in order to
engineer the required nonlinear photonic quasicrystals.  These devices
attest to the general nature of the quasicrystal-based solution to the
multiple phase-matching problem.  In Sec.~\ref{sec:cut} we show that
for collinear devices, in which all the participating waves propagate
in the same direction, a certain degree of flexibility can be obtained
by using a physical realization of the cut-and-project approach, in
which one generates the required 1-dimensional photonic quasicrystal
by cutting through a periodic crystal in 2 dimensions.

\section{Reciprocal-lattice engineering with the dual-grid method}
\label{sec:dgm}

After selecting the nonlinear medium of choice and the appropriate
operating temperature for a given frequency-conversion application,
one can calculate the required set of mismatch vectors
$\Delta\kv^{(j)}$, $j=1\ldots N$, that are to be engineered into the
reciprocal lattice of the nonlinear photonic quasicrystal. As we have
described in Ref.~\cite{Lifshitz_PRL2005}, once these mismatch vectors
are known, one can use de-Bruijn's dual grid method~\cite{deBruijn},
as generalized by G\"{a}hler and
Rhyner~\cite{GahlerRhyner_Equivalence} and later by Rabson, Ho, and
Mermin~\cite{RabsonHo1988,RabsonHo1989}, in order to design the
required quasicrystal. If the $N$ mismatch vectors are
integrally-independent, one simply uses each vector $\Delta\kv^{(j)}$,
$j=1\ldots N$, to define a family of equally-spaced parallel lines,
separated by $2\pi/|\Delta\kv^{(j)}|$, and oriented in the direction
of $\Delta\kv^{(j)}$. The set of all $N$ families constitutes the dual
grid, which is then used in the standard manner~\cite{Senechal} to
define a set of $N$ tiling vectors $\av^{(j)}$, and to calculate the
integral linear combinations of these tiling vectors that form the
vertices of the tiles in the desired quasicrystal.  The final step is
to decorate each tile with an optimal motif, {\it i.e.} to decide
which regions of the tile will be altered such that the relevant
components of the quadratic dielectric tensor $\chi^{(2)}$ are
positive, leaving the remaining background with an unchanged
negative $\chi^{(2)}$. If the $N$ mismatch vectors happen to be
linearly-dependent, one has to consider the pros and cons of selecting
a linearly-independent subset {\it vs.} using the full
linearly-dependent set for generating the dual
grid~\cite{Lifshitz_PRL2005}.

\subsection{One dimensional implementation}

\begin{figure}[tp]
\centering
\includegraphics{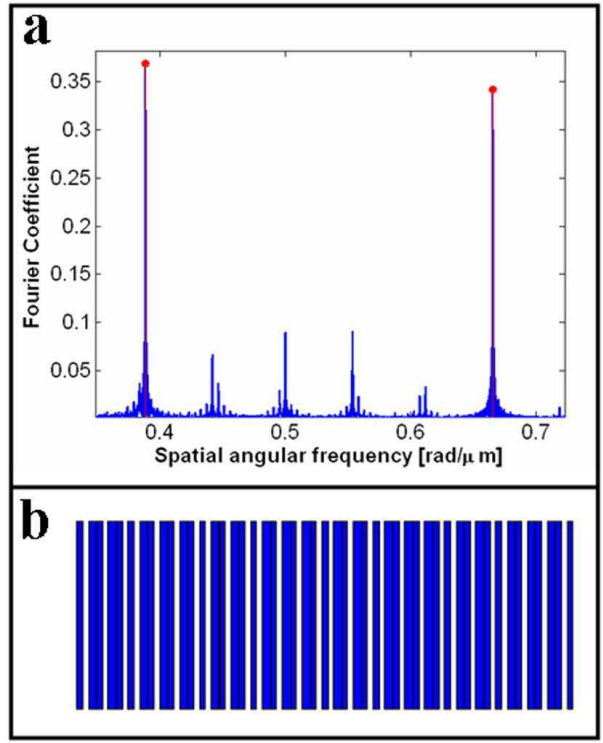}
\caption{(Color online) The polarization switch device. (a) The
  Fourier transform of the device, showing strong Bragg peaks at the
  desired mismatch vectors, $0.39\micron^{-1}$ and $0.66\micron^{-1}$.
  (b) A section of the one-dimensional nonlinear photonic
  quasicrystal. The two colors represent the two different values,
  positive and negative, of the nonlinear polarization. The smallest
  element size is $5.35 \micron$, corresponding to one half of the
  $10.7 \micron$-wide strips.}
 \label{Fig:PolRot}
\end{figure}

As a rather simple demonstration of our general solution to the
phase-matching problem, we have recently implemented a three-wave
doubler~\cite{BahabadJOSAB2007}. The reader who is interested in
implementing his or her own devices, based on our solution to
phase-matching problem, is kindly referred to this article for a
detailed pedagogical explanation of each and every step. This is a
one-dimensional device that is able to phase match three collinear
second harmonic generation processes simultaneously, taking three
input beams with wavelengths 1530nm, 1550nm, and 1570nm, and producing
three output beams at twice their frequencies. We note that the three
processes implemented in this device are independent, and therefore
could also be phase matched by fabricating a sequence of three
\emph{periodic} nonlinear photonic crystals, one for each frequency
doubling process.  Nevertheless, we were able to show that it is more
efficient to simultaneously phase match all three process in a single
\emph{quasiperiodic} nonlinear photonic crystal.

Here we implement a more elaborate one-dimensional device utilizing
our method, which is a nonlinear polarization
switch~\cite{Saltiel_PolarizationSwitching_OptLett1999}. This device
takes a beam of frequency $\omega$ and polarization $y$, propagating
in the $x$ direction, and converts some of its energy into a beam of
the same frequency, propagating in the same direction, but with
polarization $z$. It is a cascaded device~\cite{Stegman_Cascading1996}
in which the output of the first process is used as the input of the
second process. The first is a standard (type I) second harmonic
generation process with $\omega_y + \omega_y \rightarrow 2\omega_y$,
followed by a (type II) difference frequency generation process, in
which the lower-frequency waves have different polarizations,
$2\omega_y - \omega_y \rightarrow \omega_z$.

We implement the polarization switch using a LiNbO$_3$
ferroelectric crystal at a temperature of $150^\circ$
centigrade. Because linear dispersion is different for the
differently-polarized beams, the mismatch vectors for the two
processes are not equal, and are calculated using tabulated properties
of LiNbO$_3$~\cite{Jundt_LiNbO3Sellmeier_OptLett1997,%
  Edwards_LiNbO3SellmeirOQE1984} to be $0.39\micron^{-1}$ for the
first process, and $0.66\micron^{-1}$ for the second process. Using a
one-dimensional version of the dual-grid
method~\cite{BahabadJOSAB2007} we calculate the sequence of two tiling
vectors, of lengths $10.70 \micron$ and $7.63 \micron$, that generates
a one-dimensional quasicrystal with the desired wave vectors in its
reciprocal lattice. The actual one-dimensional quasicrystal is
realized by using strips of widths $10.70 \micron$ and $7.63 \micron$,
arranged in the $x$ direction according to the quasiperiodic sequence
determined by the dual-grid method. We use numerical optimization to
find the optimal duty cycles for the two strips. We find that the best
efficiencies for the desired processes are obtained when using duty
cycles of $0.5$ and $1$ for the $10.7 \micron$ and $7.63 \micron$
tiling vectors respectively. This means that one half of each $10.7
\micron$ strip is positively-poled and the other half
negatively-poled, while the narrow $7.63 \micron$ strips are
completely positively-poled. A section of the resulting
one-dimensional photonic quasicrystal is shown in
Fig.~\ref{Fig:PolRot}(b). For a single process the best efficiency is
achieved with a periodic nonlinear photonic crystal with a $0.5$ duty
cycle for its single tile. In that case the relevant Fourier
coefficient has the value of $2/\pi\simeq 0.636$. This value is
comparable to twice the Fourier coefficients at the required mismatch
vectors of the polarization-switch device, shown in
Fig.~\ref{Fig:PolRot}(a). However here we phase-match two processes
simultaneously.

\subsection{Two dimensional implementation}

\begin{figure}[tp]
\centering
\includegraphics{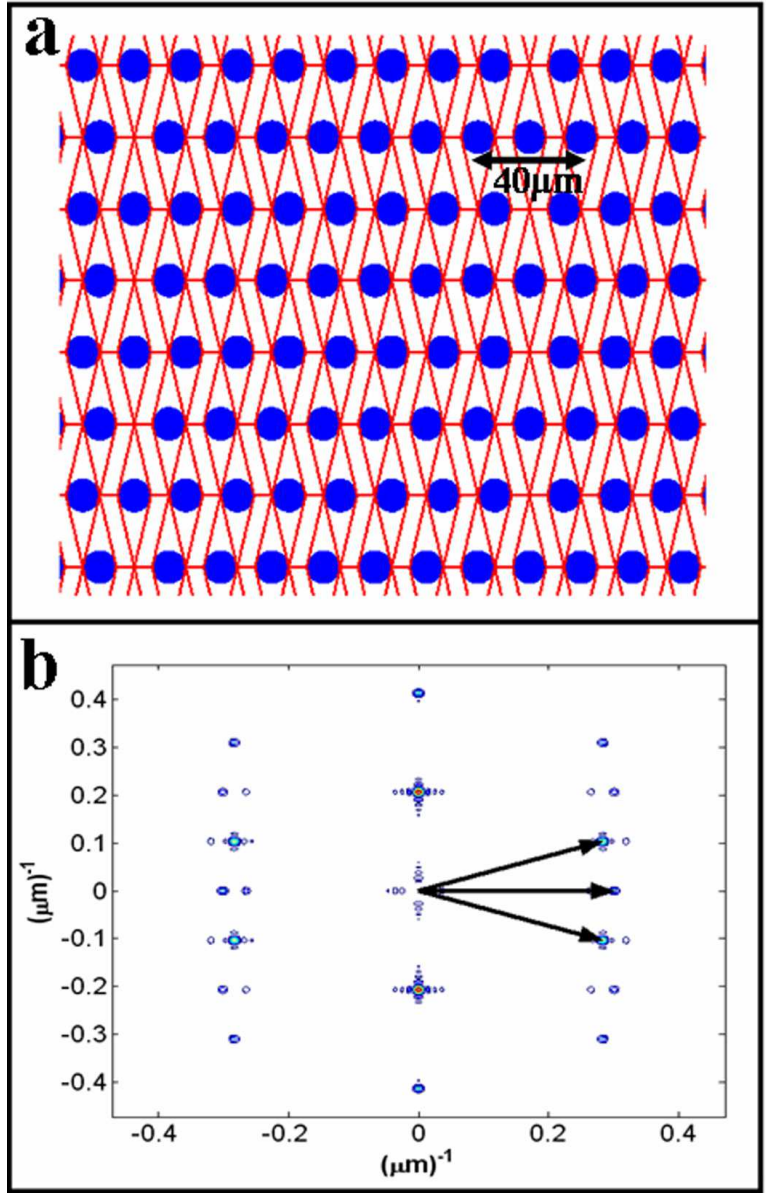}
\caption{(Color online) The multi-directional second harmonic
  generator.(a)Image of the device made of circular motifs,
  superimposed with the underlying quasiperiodic tiling. (b)
  Calculated diffraction diagram. The arrows indicate the reciprocal
  lattice vectors used to phase match the three second harmonic
  generation processes.}
 \label{Fig:SHG_fan}
\end{figure}

Next we demonstrate the versatility of our approach by designing a
multi-directional second harmonic generator. This is a two-dimensional
non-collinear device that takes three beams of wavelength 1550nm,
propagating at angles of $0^\circ$ and $\pm 20^\circ$, and generates
three output beams at the same directions at twice the input
frequency.  Using Stoichiometric LiTaO$_3$ and operating at
$100^\circ$ centigrade, the magnitude of the mismatch vectors, for all
three processes, is calculated~\cite{Bruner_SellmeirSLT2003} to be
$0.301\micron^{-1}$.  The required two-dimensional photonic
quasicrystal must clearly be symmetric with respect to the propagation
direction of the input beam.  Using a two-dimensional version of the
dual-grid method~\cite{Lifshitz_PRL2005}, we find that it is generated
by three tiling vectors, given in polar coordinates by ${\bf a}^{(1)}=
(7.54\micron, \angle 0^\circ)$, ${\bf a}^{(2)}= (31.31\micron, \angle
76.9^\circ)$, and ${\bf a}^{(3)}=(31.31\micron, \angle -76.9^\circ)$.
This yields two types of tiles---a rhombus whose edges are
31.31$\micron$ long, and a parallelogram whose edges are
31.31$\micron$ and 7.54$\micron$ long, the latter appearing in two
mirror-related orientations.

For ease of fabrication we limit ourselves to decorating the center of
each tile with a positively-polarized circle, leaving the remaining
background negatively-polarized. We employ numerical optimization only
for determining the radii of the circles for the different tiles. We
find that the best efficiencies for the desired processes, with the
largest magnitudes of the relevant Fourier coefficients, are obtained
for using a maximally-inscribed circle within the rhombic tile, and
using no circles within the parallelograms. An image of the decorated
tiling, as designed by our procedure, is shown in
Fig.~\ref{Fig:SHG_fan}(a). The calculated spectrum of the device is
shown in Fig.~\ref{Fig:SHG_fan}(b). The reciprocal lattice vectors
that phase match the three processes are indicated in this figure.
The calculated magnitudes of the Fourier coefficients for the three
mismatch vectors are $0.1$ and $0.19$ for the $0^\circ$ and $\pm
20^\circ$ processes respectively.

\goodbreak

\section{Physical realization of the cut and project method}
\label{sec:cut}

We have recently described an alternative scheme applicable to
collinear devices~\cite{Bahabad_Projection_PRL2007}, in which one
generates the required 1-dimensional photonic quasicrystal by a
process that could be thought of as a physical realization of the
cut-and-project method~\cite{Elser_Projection,DuneauKatz1985}. Here we
wish to elucidate some of its geometric features, and by doing so to
emphasize its advantage for producing tunable devices. The reader,
interested in actual implementation details of this scheme, is kindly
referred to Ref.~\cite{Bahabad_Projection_PRL2007}.  The basic idea is
to fabricate a two-dimensional periodic crystal and employ the
cut-and-project method to obtain a one-dimensional quasiperiodic
crystal, capable of phase matching two independent collinear
frequency-conversion processes. The cut is realized by taking
advantage of the fact that the input beam is not an idealized plane
wave of infinite transverse extent but actually has a finite width
$W$, for example with a Gaussian profile. Thus, the interaction of the
beam with the nonlinear medium is restricted to a strip-like region of
width $W$ along the propagation direction of the beam. Only those
lattice sites of the two-dimensional crystal that fall within this
strip in the transverse, or perpendicular ($\perp$), direction
contribute to the phase matching, and are effectively ``projected''
onto a one-dimensional quasicrystal along the propagation, or
parallel ($||$), direction.

The Fourier transform of the original 2-dimensional periodic crystal
contains Bragg peaks at wave vectors $\KV$ that form a periodic
reciprocal lattice. As we know from the cut-and-project method, the
Fourier transform of the strip-like interaction region is
1-dimensional. Each 2-dimensional Bragg peak at wave vector $\KV$
gives rise to a 1-dimensional Bragg peak at the parallel component
$\KV_{||}$ of the original wave vector, whose intensity depends on the
perpendicular component $\KV_\perp$ of the same 2-dimensional wave
vector. Because the Fourier transform of a Gaussian is also a
Gaussian, if for example the beam has a Gaussian profile and we ignore
any spreading of this profile as it propagates, then the dependence on
$\KV_\perp$ is a simple Gaussian.  Thus, given a pair of collinear
mismatch wave vectors, one simply needs to find the appropriate angle
with which to cut through the two-dimensional structure, so as to
obtain two parallel projections $\KV_{||}$ with the required mismatch
values, preferably of wave vectors $\KV$ with a small $\KV_\perp$
component.  Thus, even with a prefabricated two-dimensional periodic
crystal, one has the ability to tune the device by varying the cut
angle through the crystal, enabling the use of one device for
different combinations of frequency conversion processes.

\begin{figure}[tp]
\centering
\includegraphics{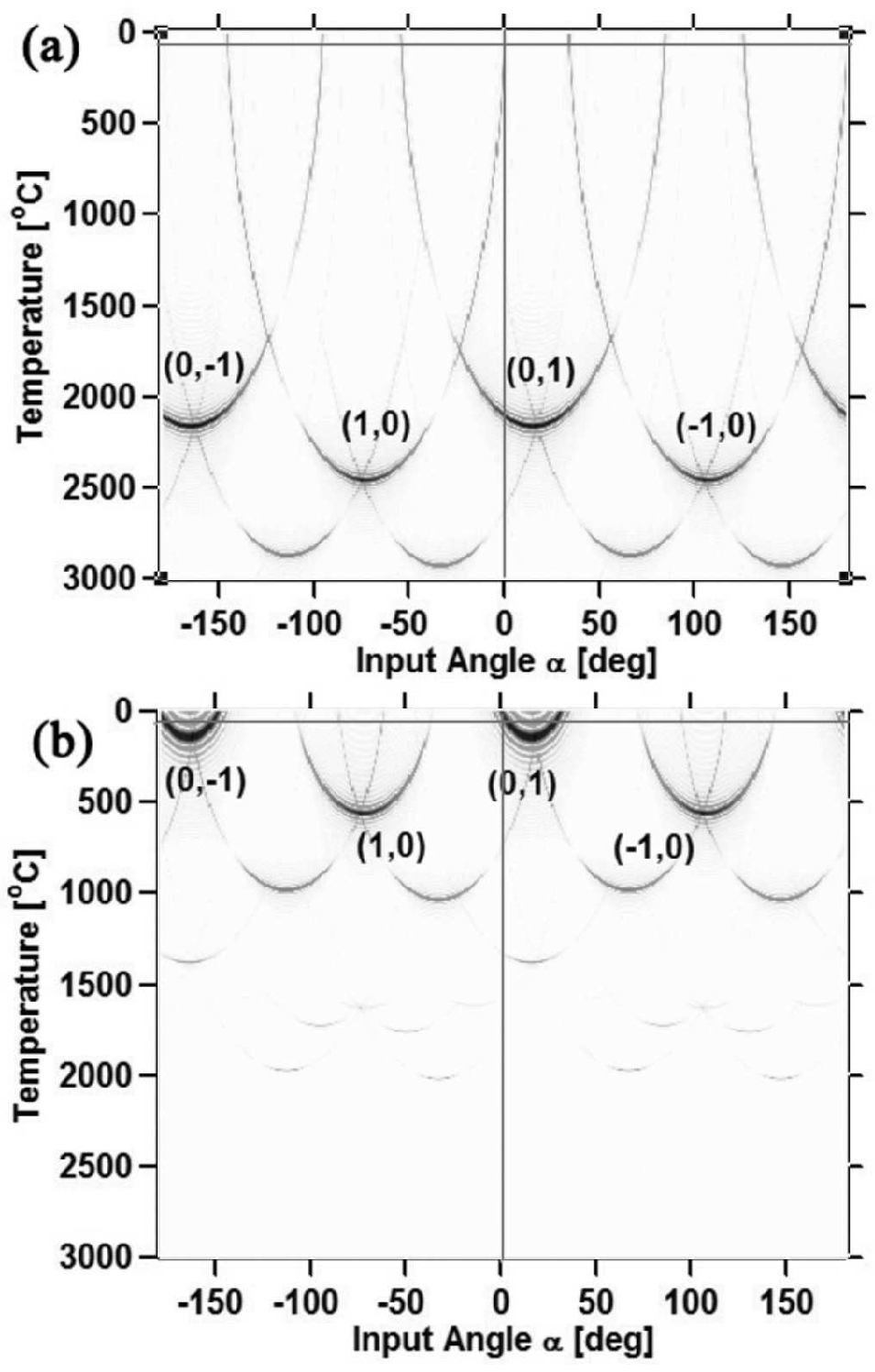}
\caption{Phase matching using the cut-and-project method.  The
  calculated combinations of temperature and propagation direction
  that satisfy the phase-matching requirements for two collinear
  second harmonic generation processes using the oblique crystal,
  described in the text.  The working point conditions are denoted by
  the intersection of the two straight lines at input angle
  $\alpha=0^\circ$ and temperature $T=23^\circ$C. The two simulated
  cases are: {\bf (a)} for an input beam of wave length
  $\lambda=1550$nm, where phase matching is realized by a projection
  of the (1,0) reciprocal lattice vector; and {\bf (b)} for an input
  beam of wave length $\lambda= 1047.5$nm, where phase matching is
  realized by a projection of the (0,1) reciprocal lattice vector.}
 \label{Fig:ProjectionFig}
\end{figure}

To illustrate how this approach can be used, we consider a nonlinear
photonic crystal built upon an oblique periodic lattice, defined by
the primitive vectors $\av^{(1)} =(6.2\micron,\angle -75^\circ)$ and
$\av^{(2)} = (7.4\micron,\angle 17^\circ)$, by associating a
positively-poled circular motif of radius $2.5\micron$ with every
lattice point. We wish to phase match two collinear second harmonic
generation processes of fundamental beams with wavelengths 1550nm and
1047.5nm. If we use stoichiometric LiTaO$_3$ and operate at room
temperature the phase mismatch values are calculated to
be~\cite{Bruner_SellmeirSLT2003} $\Delta k_1=0.297\micron^{-1}$ and $
\Delta k_2=0.820\micron^{-1}$, respectively. By varying the operating
temperature we can change the required mismatch wave vectors, and by
selecting the propagation direction of the input beams we can vary the
projected $\KV_{||}$ components. A simulation of all possible
combinations of temperature and propagation angles that satisfy the
phase matching conditions for the two processes is shown in
Fig.~\ref{Fig:ProjectionFig}. The simulation was carried out for an
interaction length of 1mm and a $20\micron$-wide square-shaped beam
profile.  Darker shades correspond to higher efficiencies. Each
parabola corresponds to a given reciprocal lattice vector $\KV$ of the
two-dimensional crystal. The apex of the parabola corresponds to
phase-matching using the whole reciprocal lattice vector
($\KV_\perp=0$), while the other parabola points correspond to
projection-based phase matching ($\KV_\perp \neq 0$). The intersection
of the straight lines is the desired working point of this device at
room temperature, where both processes are phase-matched
simultaneously.

\section*{Acknowledgment}

This research is funded by the Israel Science Foundation through grants
960/05 and 684/06. 

\bibliography{Alonbib}
\bibliographystyle{apsrev}

%\label{lastpage}

\end{document}